\documentclass[doublecol]{epl2}
\usepackage{amssymb}
\usepackage{amsmath}
\usepackage{graphicx}
\usepackage{epstopdf}

\shorttitle{Multidomain switching in the ferroelectric nanodots}
\shortauthor{F. Author \etal}
\institute{
\inst{1} University of Lorraine, IECL, CNRS UMR 7502, 54506 Vand{\oe}uvre-l\`{e}s-Nancy Cedex, France\\
\inst{2} University of Picardie, Laboratory of Condensed Matter Physics, 80039 Amiens Cedex 1, France\\
  \inst{3} L. D. Landau Institute for Theoretical Physics, Moscow, Russia } 
\pacs{77.80.Dj}{Domain structure; hysteresis} 
\pacs{64.60.an}{Finite-size systems}
\pacs{02.60.Cb}{Numerical simulation; solution of equations}
 \abstract{Controlling the polarization switching in the ferroelectric nanocrystals, nanowires and nanodots has an inherent specificity related to the emergence of  depolarization field that is associated with the spontaneous polarization. This field splits the finite-size ferroelectric sample into polarization domains.   Here, based on  3D numerical simulations,  we study the\ formation of 180$^{\circ }$ polarization domains in a nanoplatelet, made of uniaxial ferroelectric material, and show that in addition to the polarized monodomain state,
 the multidomain structures, notably of stripe and cylindrical shapes, can arise and compete during the switching process. The multibit switching protocol between these configurations may be realized by temperature and field variations.  }
\input{tcilatex}

\begin{document}

\title{Multidomain switching in the ferroelectric nanodots}
\author{Pierre-William Martelli\inst{1} \and S\'{e}raphin M. Mefire\inst{1}
\and Igor A. Luk'yanchuk\inst{2,3}}
\date{\today }
\maketitle

\begin{figure}[!b]
\centering
\includegraphics [width=5.7cm] {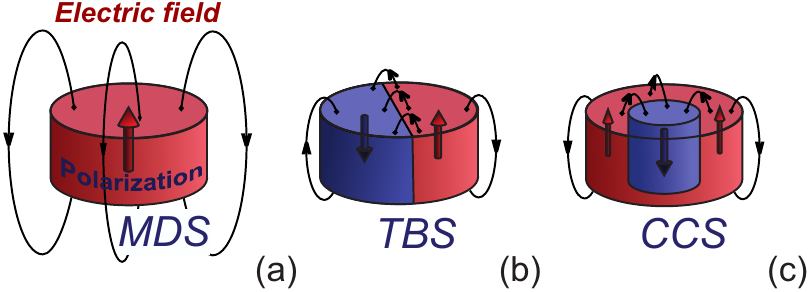} \newline
\newline
\includegraphics	[width=5cm] {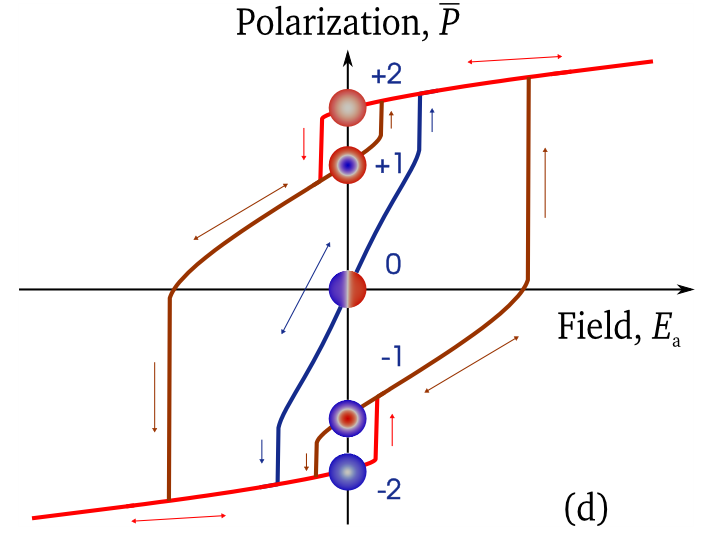}
\caption{Distribution of the depolarization electric field in the
ferroelectric nadodot (ND) with the polarized monodomain state, MDS (a), the
two-band state, TBS (b), and the concentric-cylindrical state, CCS (c).
Panel (d) shows the multibit switching between these states, denoted by
quantum numbers $0$, $\pm 1$ and $\pm 2$. The red and blue colors correspond
to the "up" and "down" polarization orientations. }
\label{FigMultibit}
\end{figure}

A fundamental property of ferroelectrics is the interplay between the
spontaneous polarization, $\mathbf{P}$, appearing due to the
symmetry-breaking off-center displacement of polar ions, and the long-range
depolarization electric field caused by the same polarization. The
depolarization field is induced by depolarizing charges distributed with
volume density $\rho =\mathrm{div}\mathbf{P}$ in regions where the
polarization is nonuniform and with surface charge density $\sigma =\mathbf{%
Pn}$ in the near-surface layer of polarization termination points (here, $%
\mathbf{n}=\left( n_{x},n_{y},n_{z}\right) $ is the unit vector, normal to
the sample surface and directed outside the sample). As illustrated in Fig.~%
\ref{FigMultibit}a for the finite-size sample, the depolarization field is
distributed in the inner space and in the surrounding outer space that costs
an additional electrostatic energy and impedes the formation of the
ferroelectric state.


The tendency to reduce the unfavorable action of the depolarization field
was noted by Landau and Kittel more than 60 years ago \cite%
{Landau1935,Landau8,Kittel1946}\ as able to lead to a polarization domain
patterning of the sample. As shown in Figs.~\ref{FigMultibit}bc, the formation
of the oppositely polarized 180$^{\circ }$ domains alternates the associated
depolarizing charge at the surface. This confines the depolarization field
closer to the surface, diminishing its energy. The price for the gain
however is the additional cost of domain wall (DW) creation. Hence, the
domain form and size are the result of a tiny balance between geometry,
electrostatic and ferroelectric properties of the system. Figs.~\ref%
{FigMultibit}bc exemplify the competition between the stripe and cylindric
domain patterns. The situation becomes even more delicate if an external
field is applied. The interaction with the field changes the domain
configuration, favoring the "up"-oriented domains with\ polarization
parallel to the field. The growth of the up-polarized domains continues up
to complete poling of the sample to the monodomain state.

The formation of Landau-Kittel domains having the structure of regular periodic
stripes in thin ferroelectric films was studied over the past decade both
experimentally \cite{Streiffer2002,Zubko2010,Zubko2012,Hruszkewycz2013} and
theoretically \cite{Bratkovsky2000,Kornev2004,Lukyanchuk2005,Lukyanchuk2009}%
. The period of the domain pattern was found to scale according to the
Landau-Kittel square root law as \cite%
{Bratkovsky2000,DeGuerville2005,Catalan2012}%
\begin{equation}
2d\simeq 2\sqrt{3.53\,\left( \widetilde{\varepsilon }/\varepsilon
_{c}\right) \text{ }\xi _{0}h},  \label{Period}
\end{equation}%
where $h$ is the film thickness, $\xi _{0}\simeq 1\,\mathrm{nm}$ is the
atomic-scale coherence length, $\widetilde{\varepsilon }=\varepsilon
_{p}+\left( \varepsilon _{c}\varepsilon _{a}\right) ^{1/2}$, $\varepsilon
_{c}$ and $\varepsilon _{a}$ are the intrinsic longitudinal and transversal
(to polarization and to stripes) dielectric constants of the ferroelectric
state and $\varepsilon _{p}$ is the dielectric constant of the surrounding
paraelectric environment. When the electric field, $E_{a}$, is applied, the
theoretical calculations predict the stretching of up-oriented domains and
the contraction of down-oriented domains. Interestingly, this induces the
oppositely-oriented coarse-grained depolarization field that finally leads
to the negative permittivity of the ferroelectric layer \cite%
{Bratkovsky2001,Lukyanchuk2014}. The behavior of the domain structure at
higher fields is little studied although one can guess that the
oppositely-oriented domain stripes will transform to the vanishing domain
droplets before the complete polarization of the system, similarly as it
happens in ferromagnetic samples \cite{Cape1971,Malozemoff1979}. Recent
numerical calculations for ferroelectric films \cite{Artemev2010}\ confirm
this hypothesis.

Much less is known concerning the properties of the Landau-Kittel 180$^{\circ }$
domains in finite-scale nanodot (ND) samples, with sizes comparable to the
predicted domain width $d$. Meanwhile, it is the application of the NDs that
is considered to be very promising in the emerging ferroelectric-based
nanoelectronics for the realization of high-density memory-storage units \cite%
{Catalan2012}. The experimental \cite{Tiedke2001,Schilling2009,Ahluwalia2013}
and theoretical \cite{Muench2009,Wang2010,Wu2014,Stepkova2014} efforts were
mostly concentrated on the pseudo-cubic perovskite ferroelectrics with eight
possible polarization orientations. The diversity of domains and of
polarization-vortex patterns was discovered, but the systematic study was
partially impeded by the complexity of the system. The electrostatic
depolarization effects are strongly coupled with ferroelastic ones, produced
by the polarization rotations and by the strain of the near-surface dead
layer \cite{Lukyanchuk2009a}.

In this letter we study how the Landau-Kittel structure of 180$^{\circ }$
polarization domains is formed in a finite-size sample. We show that the
confining electrostatic effects result in\ the various domain structures.
Field and temperature applications permit to realize the controllable 
\textit{multibit} switching between them, hence increasing the volume of the
writable information per ND.

To disjoint the depolarization effects, that are primary for the domain
formation, from the secondary lattice-deformation effects, we consider the
uniaxial ferroelectric material for which the ferroelastic coupling is
small. To be specific, we select the Sodium Nitrite, NaNO$_{2}$, \ as model
material. It has the orthorhombic symmetry, hence only two, "up" and "down",
orientations of the spontaneous polarization $P_{0}\simeq 12~\mathrm{\mu
C/cm}^{2}$ with respect to the $c$-axis. The anisotropic tensor of
dielectric constants has the principal values $\varepsilon _{a}\simeq 10$, $%
\varepsilon _{b}\simeq 5$~and $\varepsilon _{c}\simeq 12$ at low
temperatures \cite{Landolt2001,Axes}, providing the preferred orientation of
DW in the $bc$-plane.

The ferroelectric transition in\ the bulk NaNO$_{2}$ crystal occurs at $%
T_{c}\simeq 433\,\mathrm{K}$, the Curie constant, $C$, being equal to $5000\,%
\mathrm{K\ }$\cite{Landolt2001}. The characteristic value of the
depolarization field is estimated as $E_{0}=P_{0}/\varepsilon _{0}\simeq
13.5\times 10^{4}\,\mathrm{kV/cm}$ (here $\varepsilon _{0}$ is the vacuum
permittivity). We assume also that the crystal is cut in the form of
cylindrical nanoplatelet of radius $r\simeq 8\,\mathrm{nm}$ and of thickness 
$h\simeq 1.7\,\mathrm{nm}$, the spontaneous polarization being directed
along the cylinder axis. The ND is embedded into the dielectric matrix with $%
\varepsilon _{p}\simeq 90$. The high value of $\varepsilon _{p}$ is required
to reduce the depolarization energy. Otherwise the depolarization effects
will kill the ferroelectric phase, even in the multidomain state. With such
a selection of parameters, the characteristic domain width, calculated
according to Eq.~(\ref{Period}), is $d\simeq 7.1\,\mathrm{nm}$ that is
commensurate with the ND radius, $r$.

The principal results are illustrated in Fig.~\ref{FigMultibit}. At zero
applied field the ND can stay either in the polarized monodomain state
(MDS), Fig.~\ref{FigMultibit}a, or in one of the multidomain states. The
competition occurs mostly between the two-band state (TBS), Fig.~\ref%
{FigMultibit}b, \ and the concentric-cylindrical state (CCS), Fig.~\ref%
{FigMultibit}c, although other configurations are also possible. The average
polarization of the TBS, $\overline{P}$, is zero whereas the CCS and the MDS
have the nonzero polarizations, either ``up''-oriented or "down"-oriented. 
We ascribe the appropriate quantum numbers, $%
N=0,$ $\pm 1$ and $\pm 2$, to the TBS, CCS and MDS respectively, where
the sign reflects the orientation of $\overline{P}$.

The field application allows to jump among the TBS, CCS and MDS as sketched in
Fig.~\ref{FigMultibit}d. Usually, the TBS arises at zero-field cooling. When
the field, $E_{a}>0$, is applied the TBS stays reversibly-stable until it
jumps to the up-oriented MDS. Further field increase and reversal cycling
produce the hysteresis curve that, depending on the field variation
protocol, realizes the four-bit switching between the CCS with $N=\pm 1$ and
the MDS with $N=\pm 2$.\ Although the initial TBS with $N=0$ is no longer
accessible, it can again be achieved by \textit{"thermal reboot"} consisting
in the heating of the system to the paraelectric state with subsequent
zero-field cooling.\ \ The interplay between the TBS and the CCS during the
ND poling is a legacy of the already mentioned field-induced transition
between stripe and droplet domain configurations in an infinite system.

An analogous multiple-hysteresis phenomenon was recently observed in
mesoparticles of superconducting lead and explained by an irreversible decay
of magnetic vortex droplets \cite{Lukyanchuk2015}. The effect was shown to
be similar to the Rayleigh fragmentation of a charged liquid due to the
competition between the long-range Coulomb charge repulsion and the
short-range molecular attraction \cite{Rayleigh1882}. The buckling,
faceting, or even disintegration of ferroelectric domains also can be driven
by the long-range repulsion between DWs, associated with the electric
fringing fields of depolarization charges \cite{Lukyanchuk2014Facet}.

To give the further insight on the switching process we describe the
methodology of calculations. We consider the nonuniform Ginzburg-Landau
equation for the order parameter, spontaneous polarization $P$, coupled with
the electrostatic equation for the electric potential $\varphi _{f}$, in $%
\Omega _{f}$, the interior of the ND, \cite{Lukyanchuk2005,Lukyanchuk2009}: 
\begin{gather}
\left( t+\frac{P^{2}}{P_{0}^{2}}-\xi _{0a}^{2}\partial _{x}^{2}-\xi
_{0b}^{2}\partial _{y}^{2}-\xi _{0c}^{2}\partial _{z}^{2}\right)
P=-\varepsilon _{0}\kappa _{c}\partial _{z}\varphi _{f},  \notag \\
\varepsilon _{0}\left( \varepsilon _{a}\partial _{x}^{2}+\varepsilon
_{b}\partial _{y}^{2}+\varepsilon _{ic}\partial _{z}^{2}\right) \varphi
_{f}=\partial _{z}P.  \label{GL}
\end{gather}%
Here, $t=(T-T_{c})/T_{c}$ is the reduced temperature, $\kappa
_{c}=C/T_{c}\simeq 11.5$ is the displacive parameter \cite{Strukov2012}, $%
\varepsilon _{ic}\gtrsim 1$ is the small nonpolar contribution to the
Curie-Weiss divergency of $\varepsilon _{c}$, such as $\varepsilon
_{c}(t)=\varepsilon _{ic}+\kappa _{c}/t$ at $T\gtrsim T_{c}$. The
coordinates $x$, $y$ and $z$ are selected along the axes $a$, $b$ and $c$,
the corresponding coherence lengths $\xi _{0a}$, $\xi _{0b}$ and $\xi _{0c}$
are roughly equal to $1\,\mathrm{nm}$. To catch the generic features of the
switching we use the simplistic cubic form of the nonlinear term, digressing
from the more specific details of the transition. The parameter $P_{0}$ \ is
selected as spontaneous polarization at low temperatures. The electric field
and the electric displacement are given by $\mathbf{E}_{f}=-\nabla \varphi
_{f}$ and $\mathbf{D}_{f}=\varepsilon _{0}\mathbf{E}_{f}+\mathbf{P}_{f}$,
where the components of the polarization vector, $\mathbf{P}_{f}$, are
written as: $P_{f,x}=\varepsilon _{0}\left( \varepsilon _{a}-1\right) E_{f,x}
$, $P_{f,y}=\varepsilon _{0}\left( \varepsilon _{b}-1\right) E_{f,y}$ and $%
P_{f,z}=\varepsilon _{0}\left( \varepsilon _{ic}-1\right) E_{f,z}+P$.

The relations (\ref{GL}) are completed by the Poisson equation $\nabla
^{2}\varphi _{p}=0$ for the electric potential $\varphi _{p}$, outside the
ND. The corresponding electric field and displacement are expressed as $%
\mathbf{E}_{p}=-\nabla \varphi _{p}$ and $\mathbf{D}_{p}=\varepsilon
_{0}\varepsilon _{p}\mathbf{E}_{p}$.

Electrostatic rules imply the continuity of the potential, $\varphi
_{f}=\varphi _{p}$, and of the normal component of the electric
displacement, $\mathbf{D}_{f}\mathbf{n}=\mathbf{D}_{p}\mathbf{n}$. In our
notation, the latter constraint is written as $\varepsilon _{0}\varepsilon
_{ic}\mathbf{\partial }_{z}\varphi _{f}-P=\varepsilon _{0}\varepsilon _{p}%
\mathbf{\partial }_{z}\varphi _{p}$ for the upper and lower ND surfaces and
as $\left( \varepsilon _{a}n_{x}\partial _{x}+\varepsilon _{b}n_{y}\partial
_{y}\right) \varphi _{f}=\varepsilon _{p}\left( n_{x}\partial
_{x}+n_{y}\partial _{y}\right) \varphi _{p}$ for the lateral surface. For
the polarization we took the free boundary condition $\left( \xi
_{0a}^{2}n_{x}\partial _{x}+\xi _{0b}^{2}n_{y}\partial _{y}+\xi
_{0c}^{2}n_{z}\partial _{z}\right) P=0$ at the sample surface.


\begin{figure}[!b]
\centering
\includegraphics [width=5cm] {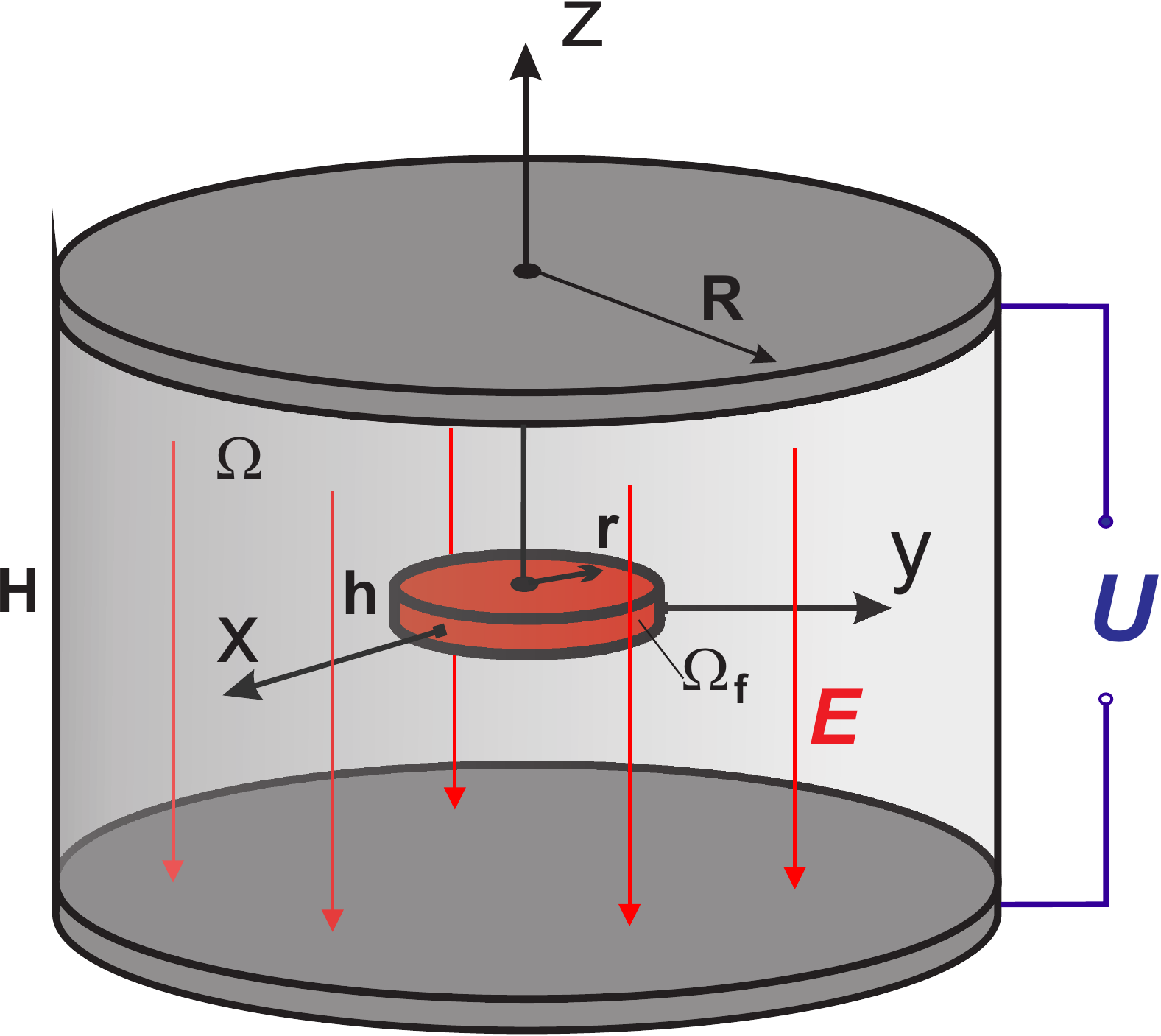}
\caption{Geometrical parameters of the system.}
\label{FigGeometry}
\end{figure}


The calculation setup is shown in Fig.~\ref{FigGeometry}. To model the outer
potential distribution, the ND was embedded into a bigger space, $\Omega $,
with sufficiently large volume to avoid to disturb the emergent fringing field.
In practice the cylinder with radius $R\simeq 12\,\mathrm{nm}$ and thickness 
$H\simeq 17\,\mathrm{nm}$ was taken, the independence of the obtained
results on the cylinder sizes was checked at each calculation stage. The
potential difference $\Delta \varphi _{p}=U$ between the top and bottom
electrodes was applied to induce the field $E_{a}=U/H$, the variation of $%
\varphi _{p}$ at the lateral surface being assumed to be linear, $\varphi
_{p}(z)=(z/H)U$.

We deal then with the boundary value problem satisfied by $P$\ and $\varphi $%
, where the electric potential in $\Omega $\ is denoted by $\varphi $, as
being $\varphi _{f}$\ or $\varphi _{p}$, namely inside or outside the ND.
The approach for solving numerically this 3D problem is based on a finite
element method, implemented as Fortran 90 homemade code (see \cite%
{MartelliPhD} for details). First, we introduced a variational formulation
of the problem, for which the variational unknowns, representing $P$ and $%
\varphi $, are found in the Sobolev spaces $H^{1}(\Omega _{f})$ and $%
H^{1}(\Omega )$ respectively \cite{Adams1975}. Then, we discretized this
formulation by making use of a mesh of $\overline{\Omega }$ (the closure of $%
\Omega $) and of the Lagrange finite elements of the first order. This mesh
consists of a collection of tetrahedra, obtained from a usual process of
triangulation, and is fine enough near the boundaries of $\Omega _{f}$ and $%
\Omega $. The discrete regions associated with $\overline{\Omega _{f}}$ and $%
\overline{\Omega }$ are polyhedral. In a correlative way, the discrete
region associated with the boundary of $\Omega _{f}$ is entirely made up of
faces of tetrahedra. Each of these faces is common both to a tetrahedron of
the discrete region associated with $\overline{\Omega _{f}}$ and another one
of the discrete region associated with $\overline{\Omega _{p}}=\overline{%
\Omega \setminus \Omega _{f}}$. By dealing with a mesh size equal to $0.47\,%
\mathrm{nm}$ for $\overline{\Omega _{f}}$ and to $0.61\,\mathrm{nm}$ for $%
\overline{\Omega _{p}}$, we were led to a square nonlinear system of 
$485621$ scalar equations. Let us mention that this system is not subject to
a uniqueness of solution, as it is the case for the associated discrete and
continuous variational formulations. Finally, we solved the system with the
help of an inexact Newton method \cite{Dennis1996}, combined at each
iteration with the Generalized Minimal RESidual (GMRES) algorithm \cite%
{Saad1986}; a preconditioner, based on an incomplete LU factorization (see,
e.g., \cite{Quarteroni2000}), was incorporated into this algorithm.

We systematically dealt with the initialization of the inexact
Newton method either with a randomly generated datum or with a solution
profile obtained at the previous calculations. The behavior of the system as
a function of the temperature or of the field was followed with the step of
at most $10^{-2}$ of the running interval whereas the finer step of $%
10^{-3}$ was used in the points of the brutal solution changes.

Eqs.~(\ref{GL}) were also studied in \cite{Nechaev2015} for the ellipsoidal
ND of triglicine sulfate (TGS), having similar dielectric properties. In
particular the temperature-induced CCS-MDS transition was observed. However,
only the axially-symmetric domain profiles (that are not always the most
stable) were obtained. For instance, the steady TBS was overlooked. This
does not permit to see all the diversity of the possible domain structures
and to investigate the field-induced switching between them, that is the
subject of the current article.

\begin{figure}[!t]
\centering
\includegraphics [width=5cm] {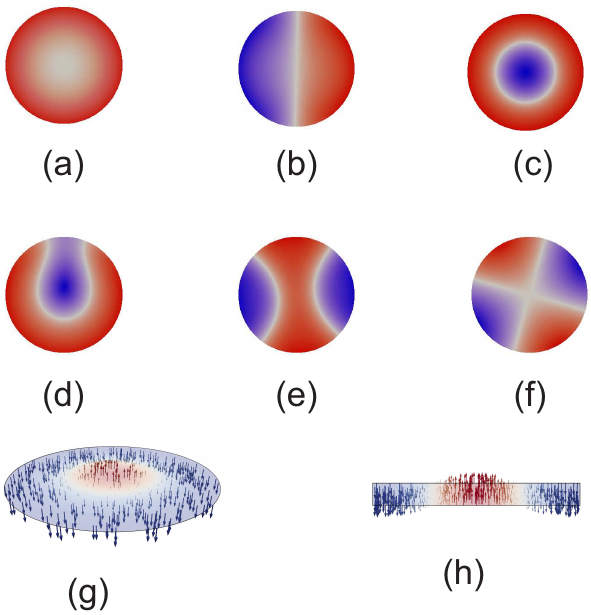}
\caption{(a)-(f) Different (meta)-stable polarization profiles, obtained
from random initial polarization distributions. Lower panel shows the
polarization distribution in the CCS for the equatorial (g) and vertical (h)
sections of the ND. }
\label{FigProfile}
\end{figure}

In what follows, we start from study of temperature evolution of the system
and show that the different \textit{metastable }domain states\textit{\ }can
be formed at the same temperature; some of them are presented in Fig.~\ref%
{FigProfile}. To catch all the possible states, we simulated the zero-field
thermal quenching to low temperatures. Random polarization and random
potential were taken at the initialization stage. Then, the final converging
state was registered. Repeating the numerical experiment with different
initial distributions, we found that, in addition to the already mentioned
MDS, TBS and CCS (Figs.~\ref{FigProfile}a-c, respectively), other profiles,
like the asymmetric-droplet state (Fig.~\ref{FigProfile}d) and the
three-band state (Fig.~\ref{FigProfile}e), can arise at $T=0$. Several other
domain patterns were also seen, but on a less regular basis. In the TBS, the
polarizations of up- and down-oriented domains are compensated, giving $%
\overline{P}=0$. In other states the average polarization is nonzero; the
latter being maximal in the MDS.

\begin{figure}[!t]
\centering
\includegraphics [width=8cm] {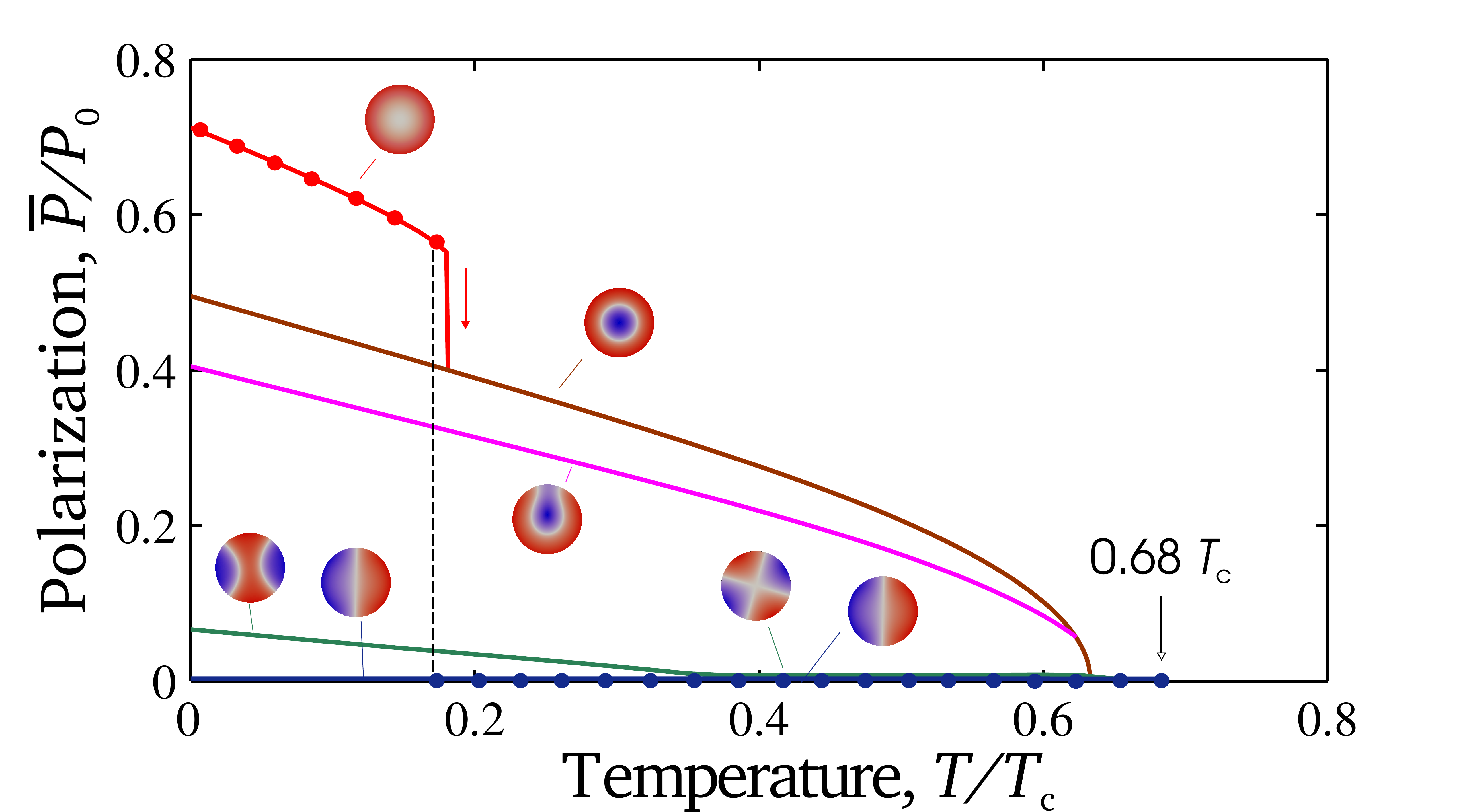}
\caption{Temperature evolution of the average polarization of the ND for
different domain configurations. Dots denote the most stable state. The
polarization is given in units of the zero-temperature bulk spontaneous
polarization, $P_0$, the temperature is given in units of the bulk critical
temperature, $T_c$.}
\label{FigTemperature}
\end{figure}

To follow the temperature evolution of the observed domain states we
performed the adiabatic heating when the next-on-heating state was obtained
from the previous one, taking the latter as the initial configuration. The
average spontaneous polarization of each of the discovered states decreases
on heating (see Fig.~\ref{FigTemperature}), each state having its own
temperature range of existence above which it irreversibly jumps to another
one, more enduring state. The corresponding transition temperature was
identified as the temperature of \textit{superheating}. Remarkably, no jump
corresponding to a \textit{supercooling} instability was observed on
adiabatic cooling.

The MDS \ stays stable on heating till \ the temperature $0.18T_{c}$, above
which it suddenly jumps to the CCS. The asymmetric-droplet state also
transforms to the CCS\ \ by almost continuous coalescence of the
edge-connecting neck but at higher temperature, $0.62T_{c}$. The average
polarization of the CCS itself gradually decreases and at $0.63T_{c}$ the
system transforms to the TBS.

The TBS is the most enduring state. Its average polarization is always zero,
whereas the maximal domain polarization also decreases on heating and
vanishes at $T_{0}=0.68T_{c}$. This temperature can be considered as the
ferroelectric transition temperature of the ND. \ As concerns the three-band
state, it remains metastable till $0.37T_{c}$. Then, it transforms to the
new diamond-like state (Fig.~\ref{FigProfile}f) with $\overline{P}=0$, which
at $0.65T_{c}$ drops to the TBS.

We identified the most stable domain configuration at each temperature,
calculating the free energy, $F=-\frac{1}{4\kappa _{c}P_{0}^{2}}\int
P^{4}d\Omega $ \cite{Lukyanchuk2009}, for each state (see the dot marks in
Fig.~\ref{FigTemperature}). The TBS remains the most stable on cooling from $%
T_{0}$ to $0.17T_{c}$ and then, the avail in energy passes to the MDS (see
dashed line in Fig.~\ref{FigTemperature}). This \textit{thermodynamic}
transition temperature is slightly lower than the given above superheating
temperature for the MDS, $0.18T_{c}$.


\begin{figure*}[!t]
\begin{center}
\includegraphics [width=13cm] {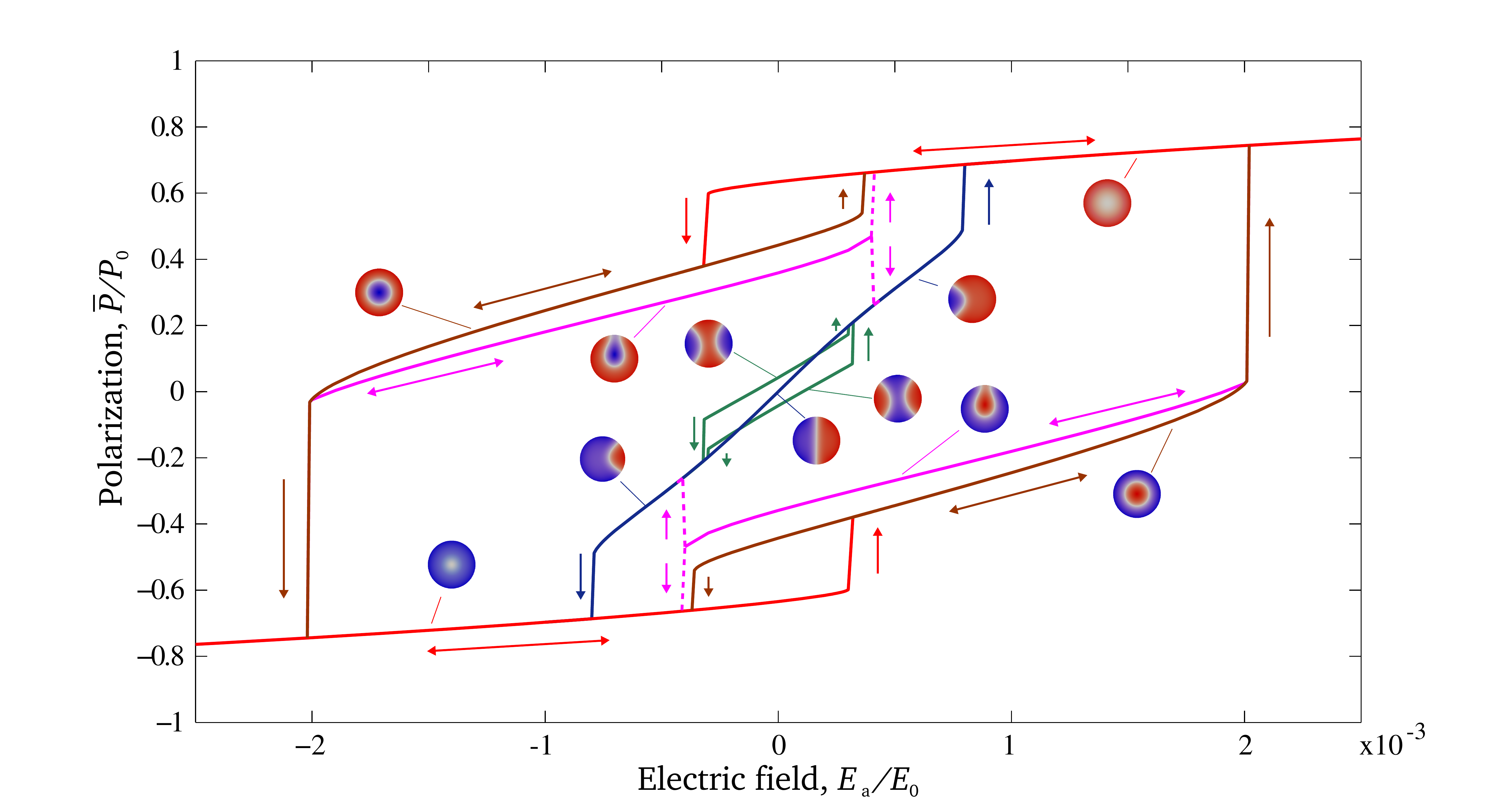}
\end{center}
\caption{Hysteresis curves for different domain states at $T=0.1T_c$. The
average polarization of the ND is given in units of the spontaneous
polarization, $P_0$ at $T=0$, the applied electric field is given in units
of the bulk depolarization field $E_0=P_0/\protect\varepsilon_0$.}
\label{FigHyst}
\end{figure*}


Figs.~\ref{FigProfile}gh illustrate the polarization distribution inside the
CCS at low temperatures. The width of the DW is of the order of $4$-$6\,\xi
_{0}$ which is larger than the typical atomistic DW width in bulk materials.
The DW has Ising-like structure when the polarization changes in 
amplitude, across the wall, but stays almost parallel to the $\mathbf{z}$%
-axis. At higher temperatures the DW\ profile becomes even softer \cite%
{Lukyanchuk2005,Lukyanchuk2009}. There is also no substantial change in the
polarization value when passing across the ND from bottom to top.\ This
makes a difference with the high-$\varepsilon $ perovskite ferroelectrics,
where the polarization flux-closure at termination points of the DW has been
observed \cite{Kornev2004,DeGuerville2005,Lukyanchuk2009}. The determining
difference is in the opposite ratio $\varepsilon _{p}/\varepsilon _{c}$ that
substantially reduces the depolarization effects in the embedded ND of NaNO$%
_{2}$. Another striking feature we observed is the deviation of DWs from the
most preferable $bc$-plane, clearly seen in the diamont-like structure (Fig.~%
\ref{FigProfile}f), and also the DWs buckling in the three-band state and,
especially, in the asymmetric-droplet state. We believe that this effect is
provided by the already mentioned interplay between the positive DW tension
energy and the negative nonlocal electrostatic energy of the fringing field 
\cite{Lukyanchuk2015}.

The behavior of the multidomain structures in external field was also studied by
a similar method of adiabatic field variation. Fig.~\ref{FigHyst} (which is
the extended version of Fig.~\ref{FigMultibit}d) presents the hysteresis
loops, $\overline{P}(E_{a})$, for various domain states at $T=0.1T_{c}$.
Monitoring the domain evolution in the applied field gives more insight into
dynamics of the switching process. Starting from the zero-field cooled TBS,
having $\overline{P}=0$, and gradually increasing the field, we follow the
central (blue) branch of the hysteresis loop. We notice that the DW in the
TBS starts to bend, to diminish the unfavorable polarization. At some
threshold field it irreversibly escapes from the sample to form the
up-polarized MDS, which is the most stable state at $E_{a}\geq 0$. The MDS 
remains with further field increase, forming the upper (red) branch of the
hysteresis loop.\ Backward field decrease keeps the MDS till $E_{a}=0$ \ and
even further, for small negative $E_{a}$, when the up-polarized MDS turns to
be metastable with respect to the down-polarized MDS.

Polarization distribution in the MDS is nonuniform. Feedback action of the
depolarization field reduces the spontaneous polarization closer to the
centers of the ND circular plane surfaces, the effect known as
Landau-Lifshitz branching \cite{Landau8}. Such a polarization central
sagging becomes especially pronounced when the field downturns below zero.
Finally, the polarization suddenly flips at the central part of the ND with
formation of the (metastable) CCS. Notably, the average polarization of the
just formed CCS\ is smaller than that in the original MDS but is still
up-oriented, that is against the applied field.

Further evolution of the system, shown by the brown line, depends on the
protocol of the field cycling. While the down-oriented field continues to
grow, the relative volume of the internal negatively-polarized cylindrical
domain also increases. At some critical field, that is nothing but the 
\textit{coercive field} of the system, the internal domain suddenly expands
to the whole volume with the formation of the stable down-polarized MDS. This
accomplishes the semi-loop of the\ global hysteresis, that can be completed
by its centrally-symmetric counterpart under the rearward field sweep. If,
however, right after the MDS$\rightarrow $CCS switching, the running of the
applied field reverses from the decrease to the increase, the internal
domain diminishes and, at some threshold $E_{a}>0$, suddenly vanishes to
restore the original MDS. This closes the local hysteresis loop that can be
observed between the MDS and CDS at small field oscillations.

The evolution of the three-band and asymmetric-droplet domain states in the
applied field (green and magenta lines in Fig.~\ref{FigHyst}, 
respectively) can be characterized as a "repulsive cycling". Once these
configurations are formed by the thermal quenching, they can live durably in
their metastable state and be locally-stable under small field variation.
However, in larger fields, they become unstable with respect to transition
to the more regular TBS, CCS or MDS. \ As soon as this happens, there is no
way to restore the original state back neither via the temperature nor via
the field variation. They disappear very soon after the regular
high-amplitude field cycling. For instance, one of the bent DWs of the
three-band state escapes on the field variation and the system jumps to the
TBS. Note the unusual crossing of the green and blue branches of the
hysteresis curve, implying that $\overline{P}$ is not a unique function of $%
E_{a}$ but depends also on the domain configuration.

The behavior of the asymmetric-droplet state is even more diverse. Depending
on the direction of the field variation, it can either join the CCS, similar
as in the temperature diagram in Fig.~\ref{FigTemperature}, or jump to the
TBS or to the MDS. The first case occurs when approaching the coercive
field in which the system drops immediately to the oppositely-polarized MDS
state. In the second case the accidental bifurcation degeneracy with respect
to selection of the switching trajectory was observed. Within the
temperature-induced fluctuation incertitude both the transitions to the TBS
or to the MDS can occur (see dashed magenta line in Fig~\ref{FigHyst}). The
interesting feature of the transition to the TBS is the lowering of $%
\overline{P}$ when the applied field increases.

The observed variety of nanoscale domain states with their irreversible
history-dependent evolution can elucidate the dynamics of the relaxor
materials that, at certain conditions, can be viewed as a set of interacting
polarization nanoclusters with intrinsic domain structure, similar to the
ferroelectric NDs. Parameters of our model system were selected in a special
way to highlight the controlled five-bit switching. Modifications of the ND
geometry and of dielectric properties of the material can lead to the
realization of other domain configurations with topologically different
hysteresis loops and even with a higher number of switching bits.

To conclude, we performed the complete analysis of the domains formation and
of the hysteresis switching process in the uniaxial ferroelectric ND as a
function of the temperature and of the applied field. Various stable and
metastable domain states were found. The most remarkable among them, the TBS,
CCS and MDS, can be formed in a controlled way, during the regular
zero-field cooling and systematic variation of the applied field. This paves
the way to the predictable individual domain manipulation on the nanoscale.
The discovered multibit switching opens new routes for the design of
high-capacity nanosize memory-storage devices in the ferroelectric-based
nanoelectronics. 

\end{document}